\documentclass[prl,twocolumn,english,superscriptaddress,floatfix,longbibliography]{revtex4-2}
\usepackage{graphicx}
\usepackage{amsmath,amssymb}
\usepackage{float}
\usepackage{xcolor}
\usepackage{appendix}

\begin{document}


\title{Engineering nonequilibrium steady states through Floquet Liouvillians
}

\author{Weijian Chen}
\email{wchen34@wustl.edu}
\affiliation{Department of Physics, Washington University, St. Louis, MO, USA, 63130}

\author{Maryam Abbasi}
\affiliation{Department of Physics, Washington University, St. Louis, MO, USA, 63130}

\author{Serra Erdamar}
\affiliation{Department of Electrical and Systems Engineering, Washington University, St. Louis, Missouri 63130}
\affiliation{Department of Physics, Washington University, St. Louis, MO, USA, 63130}

\author{Jacob Muldoon}
\affiliation{Department of Physics, Indiana University-Purdue University Indianapolis, Indianapolis, IN, USA, 46202}

\author{Yogesh N. Joglekar}
\affiliation{Department of Physics, Indiana University-Purdue University Indianapolis, Indianapolis, IN, USA, 46202}

\author{Kater W. Murch}
\email{murch@physics.wustl.edu}
\affiliation{Department of Physics, Washington University, St. Louis, MO, USA, 63130}

\date{\today}

\begin{abstract}
We experimentally study the transient dynamics of a dissipative superconducting qubit under periodic drive towards its nonequilibrium steady states. The corresponding stroboscopic evolution, given by the qubit states at times equal to integer multiples of the drive period, is determined by a (generically non-Hermitian) Floquet Liouvillian. The drive period controls both the transients across its non-Hermitian degeneracies and the resulting nonequilibrium steady states. These steady states can exhibit higher purity compared to those achieved with a constant drive. We further study the dependence of the steady states on the direction of parameter variation and relate these findings to the recent studies of dynamically encircling exceptional points. Our work provides a new approach to control non-Hermiticity in dissipative quantum systems and presents a new paradigm in quantum state preparation and stabilization.
\end{abstract}
\maketitle

Floquet engineering, which involves the control of quantum systems through periodic driving, is capable of substantially altering the energy spectra of quantum systems and inducing the formation of novel states of matter \cite{Oka2019,Regensb2012,Martin2017,Malz2021,Weitenberg2021}. One recent example is discrete time crystals observed in periodically driven many-body-localized systems \cite{Else2016,Randall2021,Frey2022}. 
For closed quantum systems, their dynamics under the periodic drive are described by the Floquet Hamiltonian \cite{Shirley1965} and can reach featureless infinite-temperature states due to the energy injection from the drive \cite{Kim2014,Lazarides2014,DAlessio2014}. However, in dissipative quantum systems, the energy injection can be consumed by dissipation, resulting in nonequilibrium steady states (NESSs) \cite{Restrepo2016,Ikeda2020}. Various interesting phenomena and applications have been demonstrated in Floquet dissipative systems, such as dissipative Floquet topological insulators \cite{Dehghani2014}, quasithermal magnetism of NESSs \cite{Schmidt2019}, and a combination of reservoir engineering with Floquet engineering \cite{Petiziol2022}.

The (stroboscopic) transient dynamics of the dissipative Floquet systems towards the NESSs under the Markovian approximation can be captured by Floquet Liouvillian superoperators \cite{gunderson2021}. These superoperators are generically non-Hermitian and therefore possess complex eigenvalues, where the real and imaginary parts determine the decay rate and the oscillation frequency of the transients. 
Recently,  
time-independent Liouvillian superoperators have drawn considerable  attention in the context of non-Hermitian physics and exceptional points (EPs) \cite{mathisen2018,Hatano2019,Minganti2019,Minganti2020,Arkhipov2020_01}, as a natural extension of Hamiltonian based non-Hermitian systems \cite{Miri2019,Ozdemir2019}. Critical damping was observed at the EPs of the Liouvillian superoperator in a dissipative qubit \cite{chen2021,chen2022}. A number of applications have been found at or near the Liouvillian EPs such as an autonomous thermal machine for entanglement generation \cite{Khandelwal2021}, quantum steering \cite{kumar2021}, and chiral Bell state transfer \cite{shishir2023}. 

\begin{figure}
\centering
 \includegraphics{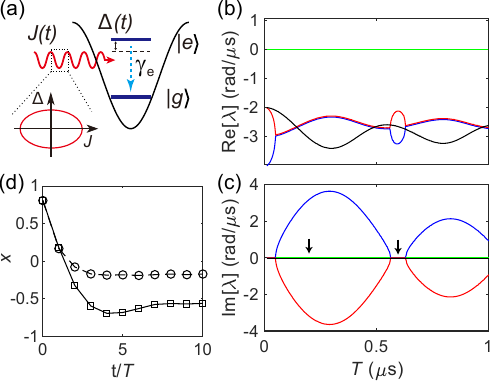} 
\caption{\textbf{Floquet Liouvillian}. (a) Schematic  of a Floquet-driven dissipative qubit, where the drive amplitude $J(t)$ and frequency detuning $\Delta(t)$ relative to the transition frequency are periodically varied. $\gamma_e$ denotes the spontaneous emission rate. Inset: One period of the Floquet drive forms a closed loop in the parameter space. (b-c) Dependence of the real (b) and imaginary (c) parts of the complex eigenvalues of the effective Floquet Liouvillian on the drive period $T$. There are four eigenvalues depicted in green, red, blue, black. Overlapping eigenvalues have been slightly offset for clarity. Parameters used for calculations: 
$J_\text{max} = 20\, \mathrm{rad\,\mu \mathrm{s}^{-1}}$,  $\Delta_\text{max} = 2\pi\, \mathrm{rad\,\mu \mathrm{s}^{-1}}$, 
and $\gamma_e=4\, \mu \mathrm{s}^{-1}$.  The eigenvalues of the effective Floquet Liouvillian describe the stroboscopic evolution. (d) Measured stroboscopic $x \equiv \langle \sigma_x \rangle$ expectation value at multiples of the loop period for $T=200$ ns (squares) and $T=600$ ns (circles). The respective underdamped and overdamped dynamics correspond to non-zero and zero imaginary parts of the eigenvalues respectively, as marked by the arrows in (c).
}
\label{fig1}
\end{figure}

\begin{figure*}[ht]
\centering
\includegraphics{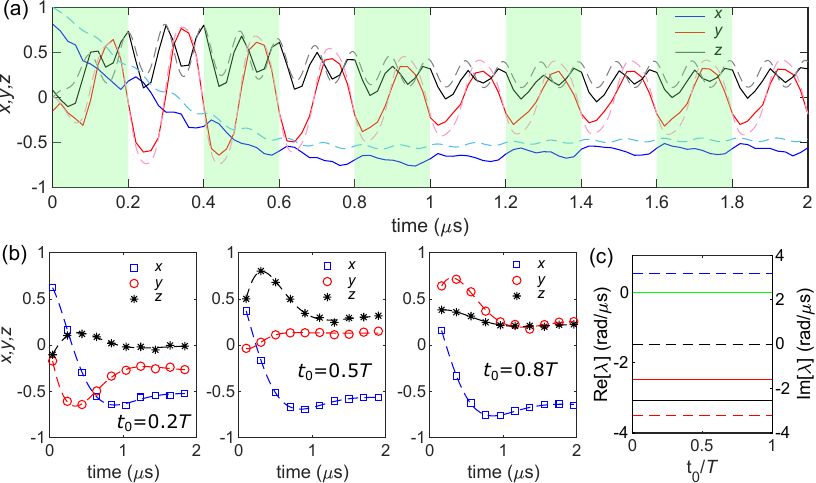} 
\caption{\textbf{Transient dynamics towards nonequilibrium steady states.} (a) Time evolution of the Pauli expectation values $x$ (blue), $y$ (red), $z$ (black).  The alternating green and white regions denote the successive drive periods of $T=200$ ns.  The initial state is $|+x\rangle$. The solid curves are experimental results, and the dashed curves are from theoretical calculations that include the effects of decay and pure dephasing. (b) Stroboscopic evolution of $x$ (blue squares), $y$ (red circles), $z$ (black stars) with different initial times $t_0=0.2T, 0.5T, 0.8T$ show different sets of Floquet evolution. The dashed curves are the fitting results. (c) Real (solid curves) and imaginary (dashed curves) parts of the complex eigenvalues of the effective Floquet Liouvillian show that they are independent of initial times $t_0$ for the same parameter path.
}
\label{fig2}
\end{figure*}

Several pioneering works have demonstrated dynamic tuning of Liouvillian superoperators, which lead to chiral state transfer \cite{chen2022} and quantum heat engines \cite{Zhang2022}. However, these works only vary the drive parameters for one period, and their connection to Floquet physics remains elusive. In this letter, we apply a Floquet drive to a dissipative superconducting qubit for multiple periods and utilize quantum state tomography to observe its transient dynamics towards the NESSs. The stroboscopic transients exhibit either overdamped, exponential decay or underdamped, oscillatory decay, where the two regimes are separated by the EPs. We use the drive period to control both the complex eigenvalues of the Floquet Liouvillians and the NESSs. Despite its strong energy dissipation, the system exhibits high-purity NESSs at certain drive periods. Finally, we connect our observations with the recent studies of chiral state transfer via dynamically tuning Liouvillians \cite{chen2022}. Our study shows that Floquet engineering provides a new resource for controlling non-Hermiticity and can help generate and stabilize quantum states.

We consider a dissipative qubit under time-dependent drive parameters, where its dynamics can be described by a Lindblad master equation:
\begin{equation}
\dot{\rho} = -i [H_c(t), \rho] + L_e \rho L_e^\dag - \frac{1}{2} \{L_e^\dag L_e, \rho\} \equiv \mathcal{L}(t) \rho.
\end{equation}
Here, $\rho$ denotes the density operator, and $L_e$ is the jump operator, defined as $L_e = \sqrt{\gamma_e} |g\rangle \langle e|$  describing spontaneous emission from level $|e\rangle$ to level $|g\rangle$ at a rate $\gamma_e$. $H_\mathrm{c}(t) = J(t) (\vert g \rangle \langle e \vert + \vert e \rangle \langle g \vert) + \Delta(t)/2 (\vert g \rangle \langle g \vert - \vert e \rangle \langle e \vert)$, characterizes coupling between the two levels by a drive with the frequency detuning $\Delta(t)$ relative to the $|g\rangle$--$|e\rangle$ transition at a rate $J(t)$. The dynamics can be fully captured by a Liouvillian superoperator $\mathcal{L}(t)$, and at each instant in time it is non-Hermitian and in general exhibits complex eigenvalues \cite{chen2022}. 

Under the condition of periodic drive, i.e., $\mathcal{L}(t) = \mathcal{L}(t+T)$, where $T$ denotes the period, it is natural to examine the dynamics in the Floquet picture. The one-period propagator for the density operator $\rho$ is given by
\begin{equation}
G(t_0,t_0+T) = \mathcal{T} e^{\int_{t_0}^{t_0+T} \mathcal{L}(t) \,dt} \equiv e^{T\mathcal{L}_{\mathrm{eff}}},   \label{eq:fp}
\end{equation}
where $t_0$ denotes the initial time, $\mathcal{T}$ denotes time-ordered integration, and $\mathcal{L}_{\mathrm{eff}}$ denotes the effective Floquet Liouvillian.  
The complex eigenvalues of $\mathcal{L}_{\mathrm{eff}}$ determine the stroboscopic evolution of the qubit state at times equal to integer multiples of the drive period (i.e., $t_0$, $t_0+T$, ..., $t_0+NT$), where $N$ represents the number of drive periods.

Inspired by the recent work  studying the dynamical tuning of Liouvillians \cite{chen2022,Zhang2022,Ding2022}, we consider a Floquet drive with both its amplitude and frequency periodically varied. 
We define the drive as $J(t) = J_\text{max} \cos(2\pi t/T)$ and $\Delta (t) = \pm \Delta_\text{max} \sin (2\pi t/T)$,
forming a closed path in the parameter space [Fig. \ref{fig1}(a)]. The signs $+$ and $-$ correspond to the counterclockwise (CCW) and clockwise (CW) directions of parameter variation, respectively. The same path is used in both our theoretical calculations and experiments. Figures \ref{fig1}(b) and \ref{fig1}(c) show the real and imaginary parts of the complex eigenvalues of the effective Floquet Liouvillian, respectively. 
The complex eigenvalues (blue and red curves in Figs.~\ref{fig1}(b) and \ref{fig1}(c)) exhibit EP transitions when varying the drive period $T$.
The eigenvalues with nonzero real parts determine the transient dynamics, and the eigenstate with zero eigenvalue corresponds to the NESS.

\begin{figure*}[ht]
\centering
\includegraphics{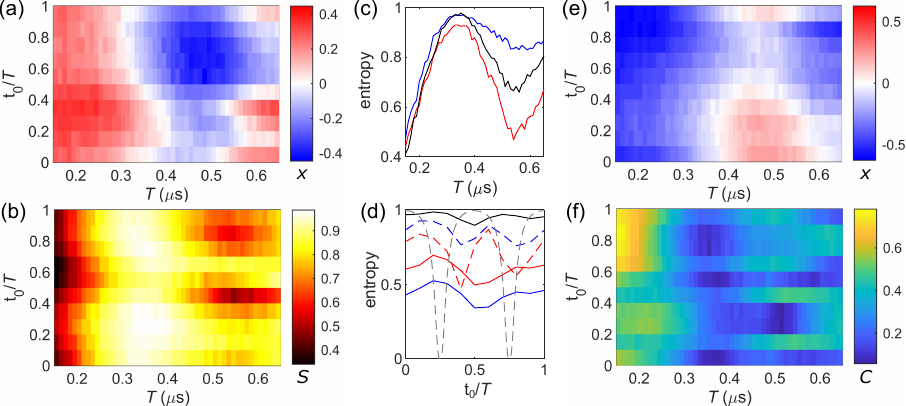} 
\caption{\textbf{Controlling nonequilibrium steady states via the drive period.} (a) The Pauli expectation value $x$ of the NESSs along the parameter path (i.e., $t_0/T$ varies between $0$ and $1$) at different periods $T$, where the parameters are varied in the CCW direction.  (b) Entropy $S$ of the NESSs calculated from the state tomography results. (c) Dependence of the entropy on the period $T$ at $t_0/T = 0.1\,\text{(blue)}, 0.4\,\text{(red)}, 0.7\,\text{(black)}$. (d) Examples of the entropy along the parameter path at different periods $T$: $150$ ns (blue solid curve), $190$ ns (red solid curve), $340$ ns (black solid curve), $440$ ns (blue dashed curve), $540$ ns (red dashed curve). For comparison, the entropy of the  
steady states of the static $\mathcal{L}(t_0)$ is also provided (dashed grey curve). 
(e) Same as (a) except that the parameters are varied in the CW direction. (f) Trace distance $C$ between the NESSs at two directions.
}
\label{fig3}
\end{figure*}

We use the lowest two energy levels ($\vert g \rangle$, $\vert e \rangle$) of a transmon superconducting circuit in this study \cite{Koch2007}. Through a reservoir engineering method, we shape the density of states of the electromagnetic field and can adjust the spontaneous emission rate of the energy level $|e\rangle$ \cite{Naghiloo2019}. In this study we set $\gamma_e \approx 4.7\,\mu \mathrm{s}^{-1}$. In addition, the qubit experiences pure dephasing  at a rate $\gamma_\phi \approx 0.3\,\mu \mathrm{s}^{-1}$. While $\gamma_\phi \ll \gamma_{e}$, the presence of this additional dephasing shifts the location of the EPs slightly, and the dynamics we observe agree well with calculations with $\gamma_e \approx 4 \ \mu\mathrm{s}^{-1}$.  The transmon circuit is dispersively coupled to a three-dimensional microwave cavity, whose resonance frequency depends on the transmon state. We can realize high-fidelity, single-shot readout of transmon state by probing the cavity with a weak microwave signal and detecting its phase shift \cite{Wallraff2005}. 

Figure~\ref{fig1}(d) provides two examples of stroboscopic evolution with the periods $T=200,600$ ns, where the parameters are varied in the CW direction. We choose the initial state $|\!+\!x\rangle$ at $t=0$.  At the end of each period, the evolution is paused, and we perform state tomography to determine the expectation values of the qubit Pauli operators
$x \equiv \langle \sigma_x \rangle$, $y \equiv \langle \sigma_y \rangle$, $z \equiv \langle \sigma_z \rangle$. 

Since this initial state does not correspond to the NESS, we expect the transient Floquet dynamics to the NESS to be characterized by the eigenvalues of $\mathcal{L}_\mathrm{eff}$. In Fig.~\ref{fig1}(d)  we display the stroboscopic evolution for $N=10$ periods.
For a $T=200$~ns period the stroboscopic evolution exhibits oscillatory decay, whereas for $T=600$~ns the evolution solely decays. 
The two types of transients correspond respectively to non-zero and zero imaginary eigenvalues of $\mathcal{L}_\mathrm{eff}$, which are separated by an EP.

Next, we perform state tomography with a finer time resolution (less than one period) to observe the full dynamics of $x$, $y$, $z$ over 10 periods of drive with $T=200$~ns [Fig.~\ref{fig2}(a)]. The evolution within one period corresponds to micromotion of the Floquet evolution. Over several periods the micromotion changes, ultimately reaching the NESS, where the micromotion repeats the same path at each period. We can quantify how the micromotion approaches the NESS by examining how the state changes from period to period. The Floquet propagator~(Eq.~\ref{eq:fp}) in fact governs an entire set of stroboscopic evolutions along the same parameter path given by different starting points $t_0\in[0,T]$. Examples of stroboscopic evolution with $t_0 = 0.2T,\ 0.5T,\ 0.8T$ are shown in Fig.~\ref{fig2}(b). Through curve fitting, we observe that the transients of $x, y, z$ with different initial time $t_0$ exhibit similar  oscillation frequencies and decay rates. 
This is because the one-period propagators with different $t_0$ have the same eigenvalues [Fig.~\ref{fig2}(c)].
Similar observations also apply to $T=600$ ns, with the full dynamics included in the Supplementary Materials \cite{supp}.
Though the one-period propagators with different $t_0$ have the same eigenvalues, the corresponding eigenstates including the one with zero eigenvalue (i.e., the NESSs) are different. This can be observed from the results in Fig.~\ref{fig2}(a), where after the fifth period the evolution is highly repetitious, meaning that it has reached the NESS: the state is still evolving in time, but the stroboscopic qubit state is now constant.

To further investigate the effect of the period $T$ on the NESSs, we use the same parameter path as given in Fig.~\ref{fig1}(a) but vary $T$ from $150$ ns to $650$ ns, and study the corresponding NESSs and their entropy at both directions of parameter variation. Here we use the results from the tenth period to approximate the NESSs. Figure \ref{fig3}(a) shows the $x$ component of the NESSs for the CCW direction. To quantify the purity of the NESSs, their entropy are also examined, defined as $S \equiv - \sum{p_i \log_2 (p_i)}$, where $p_i$ is the eigenvalue of the density matrix $\rho$ of the qubit [Fig.~\ref{fig3}(b)].
The full results of state tomography and theoretical calculations are included in \cite{supp}. As shown in Fig.~\ref{fig3}(c), the NESSs vary from high-purity states to fully mixed states by tuning the drive period.

The entropy of the NESSs along the parameter path remains relatively stable for smaller periods and increases monotonically with increasing $T$ up to $T \approx 350$ ns, where the NESSs are nearly fully mixed states [Fig.~\ref{fig3}(d)]. Further increasing $T$ lowers the entropy at a certain range of $t_0$ and leads to a double-dip feature [the red dashed curve in Fig.~\ref{fig3}(d)]. This resembles the entropy of the steady states of the (static) Liouvillian $\mathcal{L}(t_0)$ along the parameter path [the grey dashed curve in Fig.~\ref{fig3}(d)]. It minimizes when the drive amplitude approaches to zero (e.g., $t_0=0.25T$), and the steady state is close to the ground state and thus has high purity. Maximal entropy occurs when a resonant drive is applied (e.g., $t_0=0$). The quantum jumps between $|g\rangle$ and $|e\rangle$ induce transitions between the two dressed states ($|\pm x\rangle$), and the steady state is a fully mixed state \cite{Murch2012}. 

Finally, we compare the NESSs at two parameter variation directions. The $x$ component of the NESSs at the CW direction is provided in Fig.~\ref{fig3}(e), approximately opposite to that in Fig.~\ref{fig3}(a), whereas the two directions share similar $y$ and $z$ values \cite{supp}. We calculate the trace distance between the NESSs for the two directions, defined as, $C = \frac{1}{2} \mathrm{Tr}[\sqrt{(\rho_\mathrm{cw} - \rho_\mathrm{ccw})^\dag(\rho_\mathrm{cw} - \rho_\mathrm{ccw})}]$, to quantify their difference [Fig.~\ref{fig3}(f)]. Here, we reverse the time order of NESSs for one direction to calculate $C$ such that the results of two directions have a point-to-point correspondence in the parameter space. We observe that in general the parameters with high-purity NESSs overlap with those with large $C$.

\begin{figure}
\centering
\includegraphics{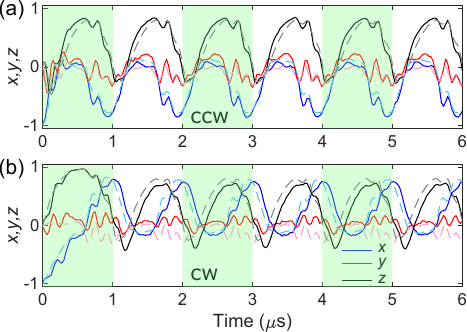} 
\caption{\textbf{Asymmetric nonequilibrium steady states at two directions of Floquet drive.} 
Quantum state tomography ($x$, blue; $y$, red; $z$, black) of dynamically tuning the Liouvillian with the period $T=1\,\mu$s. The parameters are varied in the CCW (a) and CW (b) directions. The solid curves are experimental results, and the dashed curves are from theoretical calculations.
}
\label{fig4}
\end{figure}

The observations of asymmetric NESSs at two directions in Fig.~\ref{fig3} are closely related to the recent studies of dynamically tuning the Liouvillian of dissipative quantum systems \cite{chen2022,Zhang2022,shishir2023}. 
For example, in \cite{chen2022}, we studied dynamically encircling Liouvillian EPs in the parameter space and observed chiral state transfer, where the final states are determined by the encircling directions and independent of the initial states. Such phenomena can be interpreted as non-Hermitian Hamiltonian evolution interrupted by quantum jumps between the two energy levels \cite{chen2022}. 

Here we connect those observations with the current study and show that for sufficiently large encircling time, the final states of dynamically encircling Liouvillian EPs are determined by the NESSs of the Floquet Liouvillian. We select a parameter path similar to that used in \cite{chen2022}, defined as $J(t) = (J_\text{max} - J_\text{min}) \cos^2(\pi t/T) + J_\text{min}$ and $\Delta (t) = \pm \Delta_\text{max}\sin(2\pi t/T)$, and set $T=1\,\mu$s, $J_\text{max} = 18\, \mathrm{rad\,\mu \mathrm{s}^{-1}}$, $J_\text{min} = 0.1\, \mathrm{rad\,\mu \mathrm{s}^{-1}}$, and $\Delta_\text{max} = 10\pi\, \mathrm{rad}\,\mu \mathrm{s}^{-1}$. The initial state is chosen as $|-x\rangle$.
The quantum state tomography results of six periods are provided in Fig.~\ref{fig4}. The observations in the first period are consistent with those in \cite{chen2022}: the qubit approximately returns to the initial state at the CCW direction but is transferred to $|+x\rangle$ state at the CW direction. Taking all results together, the qubit decays to the NESSs of the Floquet Liouvillian within about the first period and then evolves through the according NESSs along the parameter path. Floquet Liouvillians therefore provide a new approach to dynamical tuning of dissipative quantum systems.

\textit{Conclusion}---We have examined the transients of a Floquet driven dissipative qubit towards its NESSs. The drive period provides a means to control non-Hermiticity and NESSs. Furthermore, we have connected our observations to the recent studies in dynamical control of non-Hermitian systems governed by Liouvillian superoperators. In a broader view, our study combines Floquet engineering with reservoir engineering. Through careful design of dissipation pathways, reservoir engineering has led to stabilization of quantum states and entanglement generation \cite{Murch2012,Shankar2013, Reiter2013,Leghtas2013,Kimchi2016,Wang2023}. Combining these two techniques will provide extra degrees of freedom to create novel states for quantum technologies \cite{Petiziol2022}.

This research was supported by NSF Grant No. PHY-1752844 (CAREER), AFOSR MURI Grant No. FA9550-21-1-0202, ONR Grant No. N00014-21-1-2630, and the Institute of Materials Science and Engineering at Washington University.

%

\newpage
\pagebreak
\textcolor{white}{.}

\newpage

\widetext
\begin{center}
	\textbf{\large Supplementary materials for ``Engineering nonequilibrium steady states through Floquet Liouvillians"}
\end{center}

In the supplementary materials, we include additional experimental results and theoretical calculations for the transient dynamics and the NESSs.

\begin{figure}[ht]
\centering
\includegraphics[width=0.95\columnwidth]{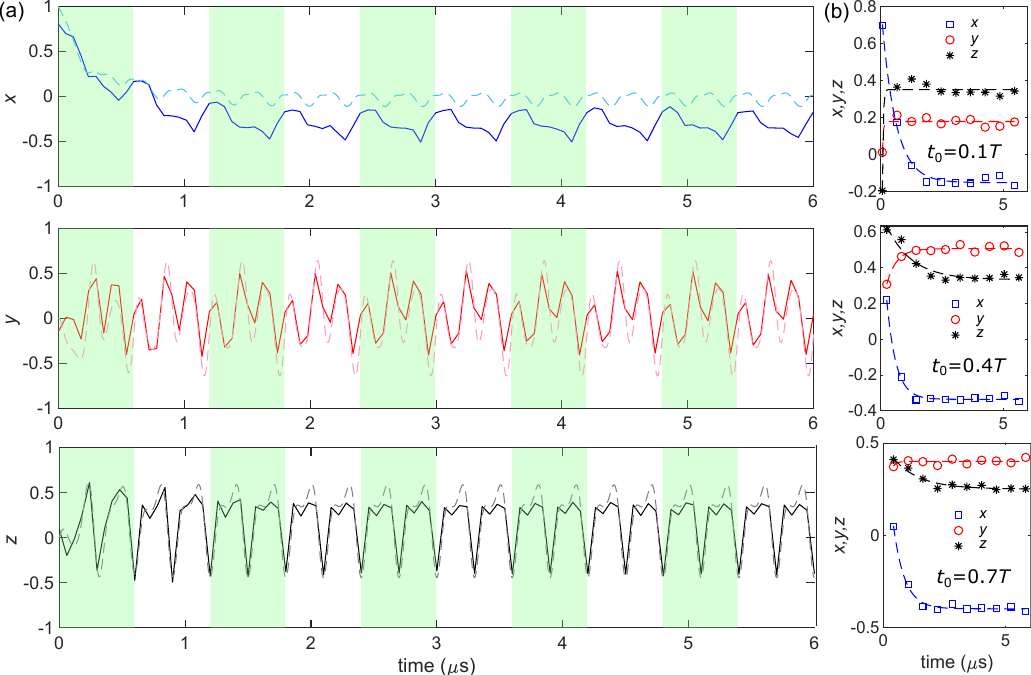} 
\caption{(a) Time evolution of the Bloch components $x$ (top), $y$ (middle), $z$ (bottom) with the period $T=600$ ns. The solid curves are experimental results, and the dashed curves are from theoretical calculations. (b)
Examples of stroboscopic evolution with $t_0=0.1T$ (top), $0.4T$ (middle), $0.7T$ (bottom). The symbols are experimental results, and the dashed curves are fitting results using exponential decay function.
}
\label{SIfig_T600ns}
\end{figure}

\begin{figure}
\centering
\includegraphics[width=0.95\columnwidth]{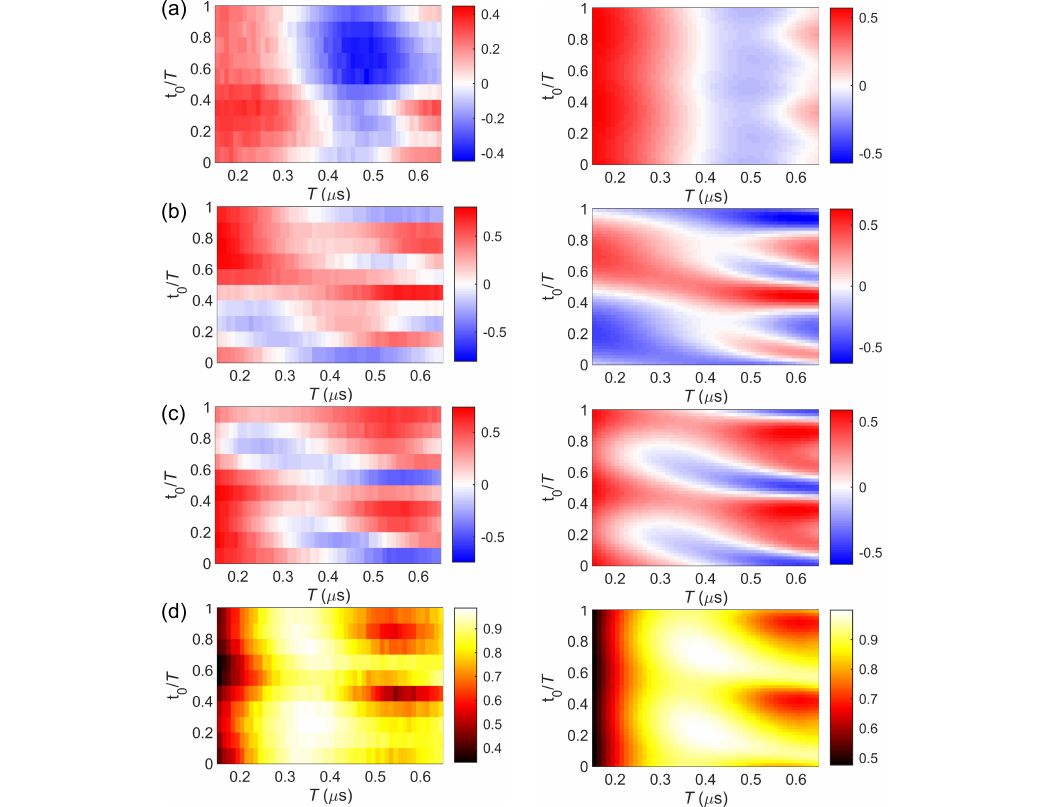} 
\caption{The Bloch components $x$ (a), $y$ (b), $z$ (c) of the NESSs and their entropy (d) along the parameter path (i.e., $t_0/T$ varies between $0$ and $1$) at different periods $T$. The parameters are varied in the CCW direction. Left: experimental results; right: theoretical calculations.
}
\label{SIfig_NESS_CCW}
\end{figure}

\begin{figure}
\centering
\includegraphics[width=0.95\columnwidth]{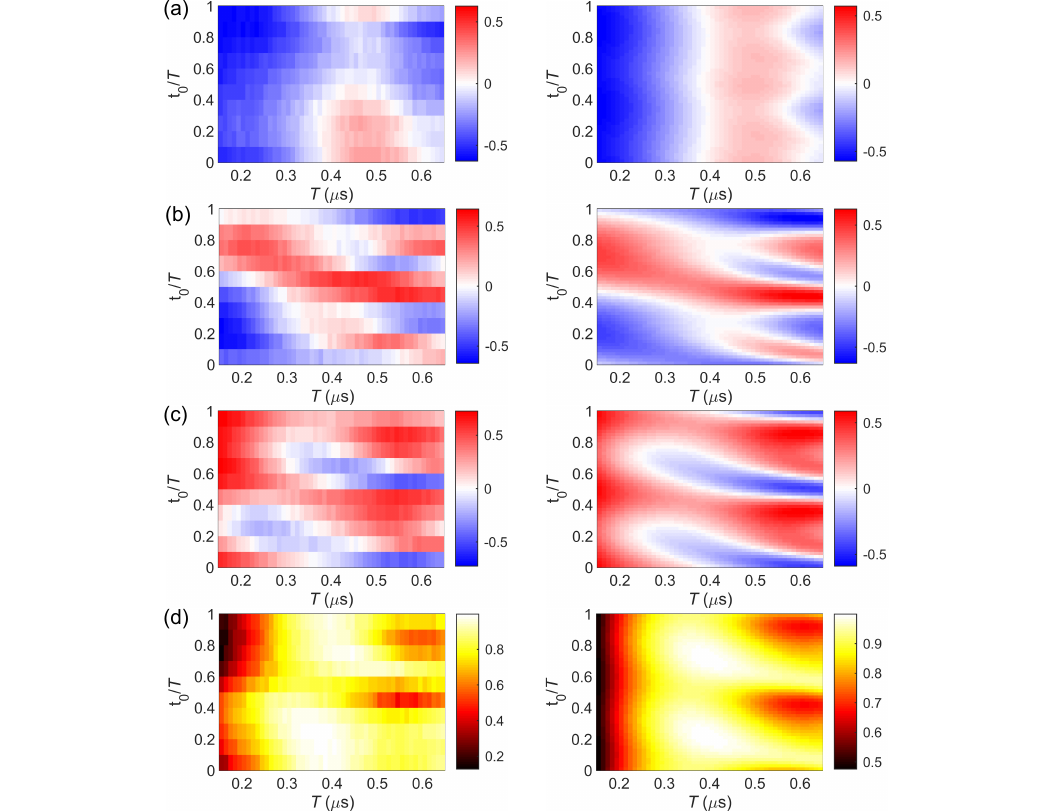} 
\caption{The Bloch components $x$ (a), $y$ (b), $z$ (c) of the NESSs and their entropy (d) along the parameter path (i.e., $t_0/T$ varies between $0$ and $1$) at different periods $T$. The parameters are varied in the CW direction. Left: experimental results; right: theoretical calculations.
}
\label{SIfig_NESS_CW}
\end{figure}

\begin{figure}
\centering
\includegraphics[width=0.95\columnwidth]{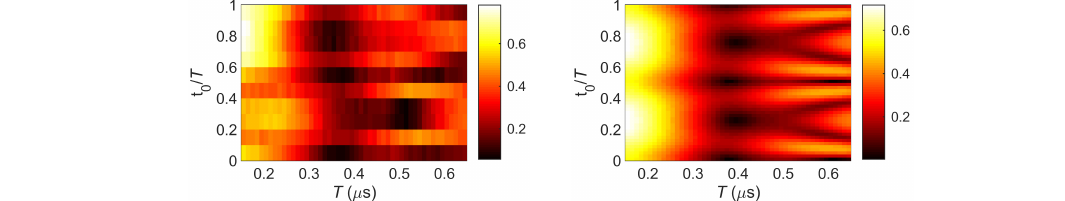} 
\caption{Trace distance between the NESSs at two directions. Left: experimental results; right: theoretical calculations.
}
\label{SIfig_NESS_chirality}
\end{figure}

Figure~\ref{SIfig_T600ns}(a) displays the full dynamics of the Bloch components $x$, $y$, $z$ within $10$ periods for $T=600$ ns. The examples of decaying stroboscopic evolution with $t_0=0.1T,\,0.4T,\,0.7T$ are shown in Fig.~\ref{SIfig_T600ns}(b). 
Figure~\ref{SIfig_NESS_CCW} shows the full state tomography results of NESSs and their entropy at varied $T$ and $t_0/T$ for the CCW direction. The experimental results [the left column in Fig.~\ref{SIfig_NESS_CCW}] are compared to the theoretical results [the right column in Fig.~\ref{SIfig_NESS_CCW}]. The results for the CW direction are provided in Fig.~\ref{SIfig_NESS_CW}. Figure~\ref{SIfig_NESS_chirality} shows the trace distance between the NESSs for the two directions, where the experimental results are also compared to theoretical calculations.

\end{document}